\documentclass[fleqn,usenatbib]{mnras}

\usepackage[T1]{fontenc}

\DeclareRobustCommand{\VAN}[3]{#2}
\let\VANthebibliography\thebibliography
\def\thebibliography{\DeclareRobustCommand{\VAN}[3]{##3}\VANthebibliography}

\usepackage{graphicx}	
\usepackage{amsmath}	
\usepackage{amssymb}	

\title[Classifier performance estimation]{Machine-learning classification of astronomical sources: estimating F1-score in the absence of ground truth}

\author[Humphrey et al.]{
A. Humphrey$^{1,2}$, W. Kuberski$^{3}$, J. Bialek$^{3}$, N. Perrakis$^{3}$, W. Cools$^{3}$, N. Nuyttens$^{3}$, H. Elakhrass$^{3}$,
\newauthor~P.~A.~C.~Cunha$^{1,4}$
\\
$^{1}$Instituto de Astrof\'{i}sica e Ci\^encias do Espa\c{c}o, Universidade do Porto, CAUP, Rua das Estrelas, Porto, 4150-762, Portugal\\
$^{2}$ DTx -- Digital Transformation CoLAB, Building 1, Azur\'em Campus, University of Minho, 4800-058 Guimar\~aes, Portugal\\
$^{3}$NannyML NV, Interleuvenlaan 62, 3001 Heverlee, Belgium\\
$^{4}$Faculdade de Ci\^{e}ncias da Universidade do Porto, Rua do Campo de Alegre, 4150-007 Porto, Portugal\\
}

\date{Accepted 2022 September 29. Received 2022 September 28; in original form 2022 July 3}

\pubyear{2022}

\begin{document}
\label{firstpage}
\pagerange{\pageref{firstpage}--\pageref{lastpage}}
\maketitle

\begin{abstract}
Machine-learning based classifiers have become indispensable in the field of astrophysics, allowing separation of 
astronomical sources into various classes, with computational efficiency suitable for application to the enormous data volumes 
that wide-area surveys now typically produce. In the standard supervised classification paradigm, a model is typically trained 
and validated using data from relatively small areas of sky, before being used to classify sources in other areas of the sky. 
However, population shifts between the training examples and the sources to be classified can lead to `silent' degradation in 
model performance, which can be challenging to identify when the ground-truth is not available. In this Letter, we present a
novel methodology using the NannyML Confidence-Based Performance Estimation (CBPE) method to predict classifier F1-score in the presence 
of population shifts, but without ground-truth labels. We apply CBPE to the selection of quasars with decision-tree ensemble models, using 
broad-band photometry, and show that the F1-scores are predicted remarkably well (${\rm MAPE} \sim 10\%$, $R^2 = 0.74-0.92$). 
We discuss potential use-cases in the domain of astronomy, including machine-learning model and/or hyperparameter selection, 
and evaluation of the suitability of training datasets for a particular classification problem. 
 
\end{abstract}




\section{Introduction}

Classification of sources is a fundamental activity in modern astronomy, allowing objects with specific properties, or in
a particular evolutionary phase, to be selected for further study. While colour-colour methods have traditionally been 
employed for object classification \citep[e.g.,][]{Haro1956,Daddi2004,Leja2019,Bisigello2020}, template-fitting methods 
have also become widely used for more detailed classification and characterisation 
\citep[e.g.,][]{Arnouts1999,Bolzonella2000,daCunha2008,Ilbert2006,Laigle2016,Gomes2017}.

In recent years, supervised machine learning has seen an explosion in popularity as a tool to classify sources into 
phenomenological types \citep[e.g.,][]{Cavuoti2014,Bai2019,clarke2020,Cunha2022}, or into morphological classes 
\citep[e.g.,][]{Dieleman2015,Huertas-Company2015,DominguezSanchez2018,Tuccillo2018,Nolte2019,Bowles2021,Bretonniere2021}.
The supervised learning paradigm allows the efficient creation of prediction functions that can be dramatically more 
scalable than template-fitting methods, while often outperforming the traditional colour-colour or template-fitting 
methods \citep{Humphrey2022}.

One potential problem with the standard supervised learning paradigm is the often tacit assumption that there is no 
population shift (also known as 'data drift') between the training samples and the samples to be classified, and that model 
performance will thus not differ between the two samples. In layman's terms, when this assumption is true, it means there is
no significant statistical difference between the samples. However, in real-world applications of supervised learning, 
population shift is frequently present and can cause `silent failure', where model predictions are significantly reduced 
in quality compared to the original validation results \citep[e.g.,][]{ElHay2022,Bennett2022}. 

The impact of population shift on machine-learning models in astronomy has received relatively little attention to date 
\citep[see, e.g.,][]{Vilalta2019,Humphrey2022}. However, population shift is likely to become increasingly relevant when supervised 
machine learning methods are used to classify, or to estimate redshifts and properties, of the billions of sources that will be detected in 
surveys such as the Euclid Wide Survey \citep{Laureijs2011,Scaramella2021}, or the Vera C. Rubin Observatory Legacy Survey of Space 
and Time \citep[LSST:][]{Ivezic2019}. For example, extragalactic deep-fields offer rich multiwavelength data suitable for accurate labelling 
of training data \citep[e.g., COSMOS:][]{Scoville2007,Laigle2016}, but their relatively small area may result in population shift with respect 
to other fields, due to cosmic or sample variance, and/or due to in-built biases (e.g., bright foregound objects may be under-represented by design). 
Therefore, it is crucial to identify methods to detect and overcome the impact of population shift on supervised machine learning in 
astronomy.

Various previous methods exist for the detection of population shifts, or to assess prediction quality
\citep[e.g.,][]{SinghSethi2017,Jiang2018,White2019,Malinin2020,Angelopoulos2022}. However, the shifts detected by these methods 
do not necessarily correlate with changes in model performance, since population shift does not always lead to a change in model performance; 
furthermore, reliable estimates of model performance, in terms of classification metric values, are not provided. Thus, there is a critical 
need for a methodology that can reliably estimate metrics of machine learning model performance, when the test-set `ground-truth' labels are 
not available, allowing model failure to be identified and quantified, and allowing also decisions to be taken concerning the suitability of 
a model or training dataset for a particular classification problem. 

In this Letter, we explore the application of the \texttt{NannyML} Confidence-Based Performance Estimation (CBPE) method for the estimation of classification
model performance, in the context of the classification of astronomical sources. This Letter is structured as follows. In $\S$\ref{metrics}, we define 
the performance metrics used. Next, in $\S$\ref{cbpe} we describe the CBPE methodology for estimating model performance in the absence of ground-truth. 
We describe in $\S$\ref{dataset} the construction of the dataset used in this study, In $\S$\ref{performance}, we explore the application of CBPE to the 
identification of quasars from broad band photometry. We briefly discuss the results and draw conclusions in $\S$\ref{conclusions}.

\section{Metric definitions}
\label{metrics}

\subsection{Classification}
To evaluate our classification results, we use the F1-score metric \citep{Dice1945,Sorensen1948}, which is the harmonic mean of the precision and recall.
The F1-score is calculated as:

\begin{equation}
    {\rm F1} = \frac{{\rm TP}}{{\rm TP} + 0.5({\rm FP} + {\rm FN})}\,,
	\label{eq:F1}
\end{equation}

\noindent where TP is the number of true positives, FP is the number of false positives, and FN is the number of false negatives. In 
addition, the number of true negatives will be referred to by the abbreviation TN. The F1-score 
can have values between 0 and 1, with 1 being the best score. 

\subsection{Regression}

To evaluate the quality of F1-score estimates, we use two regression metrics. The coefficient of determination $R^2$ is 
calculated as

\begin{equation}
    R^2 = 1 - \frac{\sum\limits_{i=1}^{n}(\hat{y_i}-y_i)^2}{\sum\limits_{i=1}^{n}(y_i-\bar{y})^2}\,,
	\label{eq:r2}
\end{equation}

\noindent where $y_i$ is the true value of the F1-score of the $i$-th sample, $\bar{y}$ is $y_i$ averaged over $n$ samples, and $\hat{y_i}$ is the predicted value. 
Higher values of $R^2$ indicate a better model, up to a maximum value of 1, with no minimum value. 
The $R^2$ score indicates how much of the variance in $y$ is explained by the independent variables present in the model. It is both scale and dataset dependent, 
which means it is often not possible to compare $R^2$ scores between different datasets, or even different subsets of the same dataset.

In addition, we use the mean absolute fractional error (MAFE), which is calculated as:

\begin{equation}
    {\rm MAFE} = \frac{1}{n} \sum\limits_{i=1}^{n} \left| \frac{\hat{y_i}-y_i}{y_i} \right| \,.
	\label{eq:mape}
\end{equation}

MAFE is scale independent, since it provides a measure of relative error. Note that MAFE is identical to the more commonly used mean absolute 
percentage error (MAPE) divided by 100.

\section{Confidence-based performance estimation}
\label{cbpe}

\subsection{Calculation method}

The CBPE method for binary classification is part of the open-source 
\texttt{NannyML}\footnote{\href{https://github.com/NannyML/nannyml}{https://github.com/NannyML/nannyml}} library for \texttt{Python}. 
In brief, CBPE uses the confidence (or lack thereof) of a classifier in its ability to correctly assign the correct class to test-set examples, 
taking advantage of the fact that the expected quality of the classification is encoded within the class probabilities that are predicted 
by the model. The method takes calibrated binary class probabilities as input, and predicts the confusion matrix 
(i.e., ${\rm \widehat{TP}}$, ${\rm \widehat{TN}}$, ${\rm \widehat{FP}}$, and ${\rm \widehat{FN}}$), as:

\begin{equation}
    {\rm \widehat{TP}} = \sum{ \{\, \hat{y} \,|\, \hat{y} \ge t \,\}\ }\,,
	\label{eq:TP}
\end{equation}

\begin{equation}
    {\rm \widehat{TN}} = \sum{ \{\, 1-\hat{y} \,|\, \hat{y} < t \,\}\ }\,,
	\label{eq:TN}
\end{equation}

\begin{equation}
    {\rm \widehat{FP}} = \sum{ \{\, 1-\hat{y} \,|\, \hat{y} \ge t \,\}\ }\,,
	\label{eq:FP}
\end{equation}

\begin{equation}
    {\rm \widehat{FN}} = \sum{ \{\, \hat{y} \,|\, \hat{y} < t \,\}\ }\,,
	\label{eq:FN}
\end{equation}

\noindent where $\hat{y}$ is the calibrated probability given by the model of an example being in class 1, and $t$ is the probability threshold used to define the 
boundary between class 0 and class 1. Equations ~\ref{eq:TP}-\ref{eq:FN} can be summarized as summations of the predicted class probabilities at and above,
or below, the probability theshold $t$.

Subsequently, the F1-score is predicted using equation~\ref{eq:F1}. While it is possible to predict other metrics, such as precision, recall, specificity, etc., 
in the interest of simplicity we restrict the present study to F1-score. 

\subsection{Assumptions}
\label{assumptions}

The CBPE method relies on frequentist inference to estimate model performance, with the following main assumptions. Firstly, it is assumed that the model scores used as input are calibrated probabilities. While classification algorithms generally provide class scores in the range 0-1, these scores often cannot be treated as class probabilities because the training process usually prioritises the minimization of a loss function, with other considerations (i.e., probability calibration) being typically less important. This is true for the learning algorithms used herein, and many other commonly used machine-learning classification methods. Thus, if a classifier outputs scores that are not calibrated class probabilities, then calibration must be performed prior to the application of the CBPE method.

Second, it is assumed that concept 
drift\footnote{We define `concept drift' as a change in the mapping between the features and the labels, such that two datasets with statistically identical features
 have statistically different labels. This is in contrast to population shift (or `data drift'), where the statistical properties of the features change without 
 changing the mapping between the features and labels. These terms are sometimes, erroneously, used interchangeably in the literature.} \citep[see, e.g.,][]{Bayram2022}
does not take place between the training data and the data subset for which the model performance is to be estimated (i.e., the test set). When concept drift is present, the CBPE method may underestimate changes in model performance, since it only estimates the performance change due to population shift. Indeed, our preliminary experiments with other real-world datasets have shown that CBPE demonstrates reduced predictive accuracy when concept drift is present.

\section{The dataset}
\label{dataset}
The sample used for this study was selected from the Sloan Digital Sky Survey \citep[SDSS;][]{Gunn1998}, using an \texttt{SQL} query via 
the SDSS SciServer CasJobs system. We selected the top 10,000,000 sources in DR15 meeting the criteria \texttt{mode = 1}, \texttt{sciencePrimary = 1}, 
and \texttt{clean = 1}. Sources with any missing photometry measurements were removed. The data were subsequently undersampled 
such that 9 in 10 rows were dropped, resulting in a dataset containing 351,089 sources (209,266 galaxes; 81,536 stars; 60,287 QSOs), 
allowing efficient model training on a `typical' laptop computer. The original dataset, along with all scripts used in this work, are publicly available 
on GitHub\footnote{\href{https://github.com/humphrey-and-the-machine/f1\_prediction}{https://github.com/humphrey-and-the-machine/f1\_prediction}.}.

Rudimentary feature engineering was performed using the \texttt{modelMag\_u}, \texttt{modelMag\_g}, \texttt{modelMag\_r}, \texttt{modelMag\_i}, \texttt{modelMag\_z}, 
and \texttt{fiberMag\_r} photometry measurements. First, we calculated all unique \texttt{modelMag} colour combinations (e.g., $u-g$ ... $i-z$). To obtain a 
feature with sensitivity to source size, we also computed \texttt{modelMag\_r}-\texttt{fiberMag\_r}. These quantities, together with the $u$,$g$,$r$,$i$,$z$ values 
of \texttt{fiberMag}, \texttt{petroMag}, \texttt{modelMag}, and \texttt{psfMag}, were used as the input features for model training (31 in total). 
Finally, the dataset was sorted in order of ascending right ascension (RA).

The classification problem considered herein is the selection of quasars using broad band photometry. Thus, the target variable is a binarized version of 
\texttt{class}, where QSOs are assigned binary class 1, and stars or galaxies are assigned 0. 

\begin{figure}
	\includegraphics[width=\columnwidth]{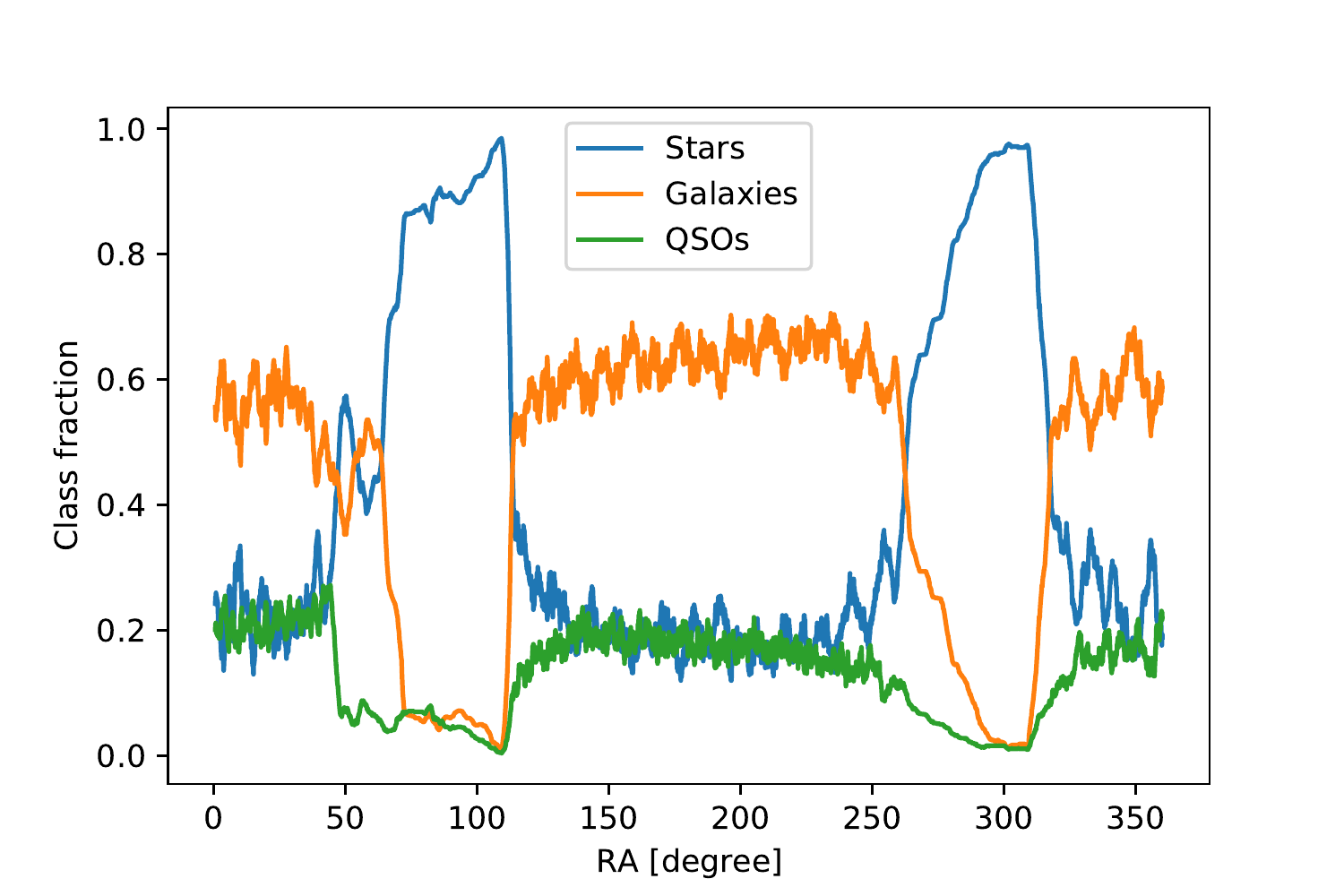}
	\includegraphics[width=\columnwidth]{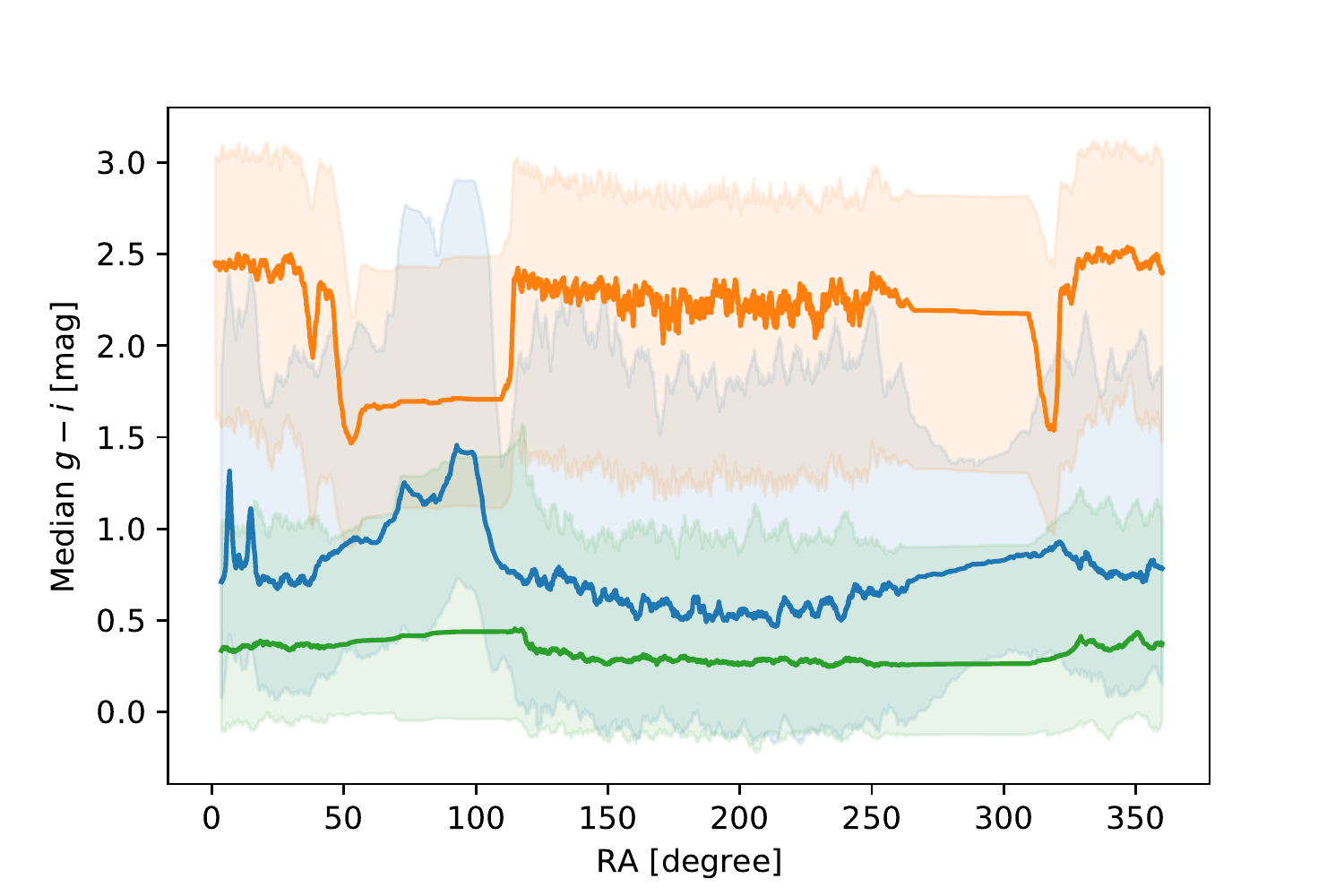}
    \caption{General characteristics of the dataset as a function of RA. {\bf Top:} The fraction of sources in each of the three classes, versus RA. 
    {\bf Bottom:} The median $g-i$ colour of each class versus RA. The shaded areas show the $\pm 1\,\sigma$ range in $g-i$ for each of the three classes. 
    In both cases, each data point represents a rolling window with a length of 1000 objects,
    and a step size of 1 object.}
   \label{fig1}
\end{figure}

Fig.~\ref{fig1} shows some of the general characteristics of the sample. As expected, there are substantial population shifts, with the balance between stars, 
galaxies, and quasars showing a strong evolution with RA (Fig.~\ref{fig1}, top panel). In particular, the number of stars relative to other sources is dramatically
higher in the ranges $50 \la {\rm RA} \la 110$ deg and $250 \la {\rm RA} \la 340$ deg, due to the presence of the Milky Way. Significant shifts are also apparent in the 
colours of the three different classes of source (Fig.~\ref{fig1}, lower panel). 

\begin{table}
	\centering
	\caption{Hyperparameters used to train the classification models. Parameters not listed here kept their default values.}

	\begin{tabular}{lcc} 
		\hline
		Learning algorithm    & Hyperparameter             & Value \\
		\hline
		\texttt{RandomForest} & \texttt{n\_estimators}      & 100 \\
		                      & \texttt{max\_depth}         & 4 \\
        \hline
		\texttt{XGBoost}      & \texttt{n\_estimators}      & 300 \\
							  & \texttt{max\_depth}         & 4 \\
						      & \texttt{tree\_method}       & hist \\
							  & \texttt{eval\_metric}       & map \\
        \hline
		\texttt{LightGBM}     & \texttt{n\_estimators}     & 1000 \\
							  & \texttt{max\_depth}        & 10 \\ 
                              & \texttt{learning\_rate}    & 0.01 \\
                              & \texttt{colsample\_bytree} & 0.9 \\
                              & \texttt{subsample}        & 0.8 \\
        \hline
	\end{tabular}
	\label{tab:global_selection}
\end{table}

\section{Model performance prediction}
\label{performance}

\subsection{Model training}
\label{training}
Models were generated using three different machine learning algorithms: the popular \texttt{RandomForestClassifier} method \citep{Breiman2001} 
from the \texttt{Scikit-Learn} package \citep{Pedregosa2011}, and the gradient-boosting decision tree methods 
 \texttt{XGBoostClassifier}\footnote{\href{https://xgboost.readthedocs.io}{https://xgboost.readthedocs.io}} \citep{Chen2016}, and 
 \texttt{LightGBMClassfier}\footnote{\href{https://lightgbm.readthedocs.io}{https://lightgbm.readthedocs.io}} \citep{Ke2017}.

Since objective of this study is to explore the application of CBPE, rather than to identify optimal learning algorithms or hyperparameters for the 
selection of quasars, the most important hyperparameters were tuned manually and non-exhaustively, with the objective of producing classification 
models that are of an acceptably high-quality. A summary of the hyperparameters used for model training is given in Table ~\ref{tab:global_selection}.

The models were trained on 1/40 of the sources in the range $150 < {\rm RA} < 200$ deg. Sources used for model training were not used 
at any subsequent stage in the workflow. 

\subsection{Probability calibration}

The aim of calibration is to derive and apply a transformation to class probabilities such that the transformed values correspond to the 
true probability of belonging to the class in question. For instance, objects assigned a class probability of 0.80 should have an 80 per cent probability 
of belonging to class 1. However, class probabilities produced by many classifiers are biased and poorly calibrated. Thus, as discussed in $\S$\ref{assumptions}, 
class probabilities must be calibrated before the CBPE is applied. 

While there are various different methods to calibrate class probabilities \citep[see, e.g.,][]{Niculescu-Mizil2005,Kull2017}, we have used the isotonic regression 
method, which is appropriate for relatively large datasets (\ga 1000 examples). This method fits a non-parametric, isotonic regressor, which yields a step wise 
non-decreasing function. 

In the interest of simplicity and reproducibility, we have used the \texttt{Scikit-Learn} function \texttt{CalibratedClassifierCV}, which produces a calibration 
transformation simultaneously with the training of the classifier, using 5-fold cross-validation. Note that the same samples cannot be used to train the 
classifier and compute the calibration transform.

\subsection{Results}

\begin{figure}
	\includegraphics[width=\columnwidth]{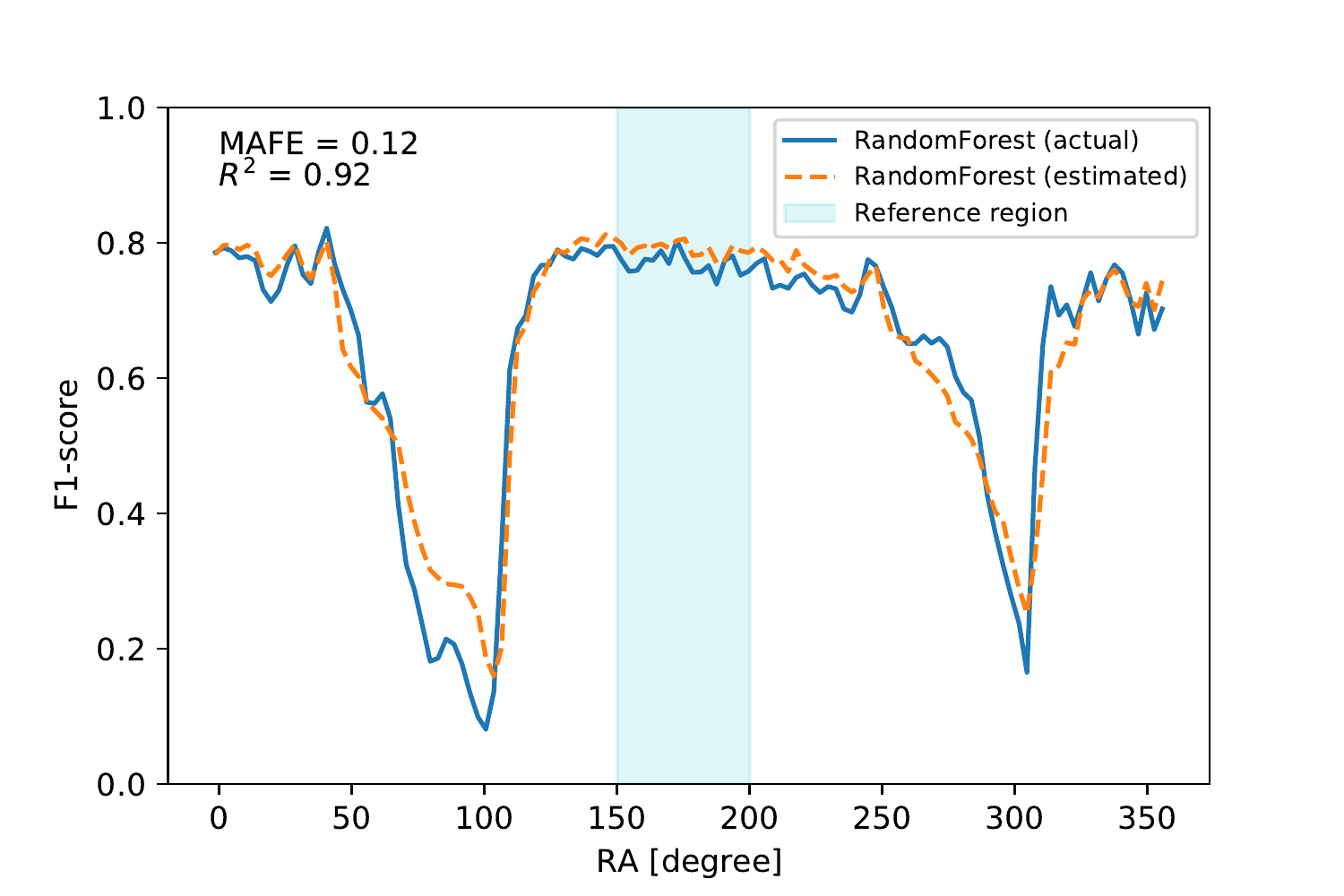}
	\includegraphics[width=\columnwidth]{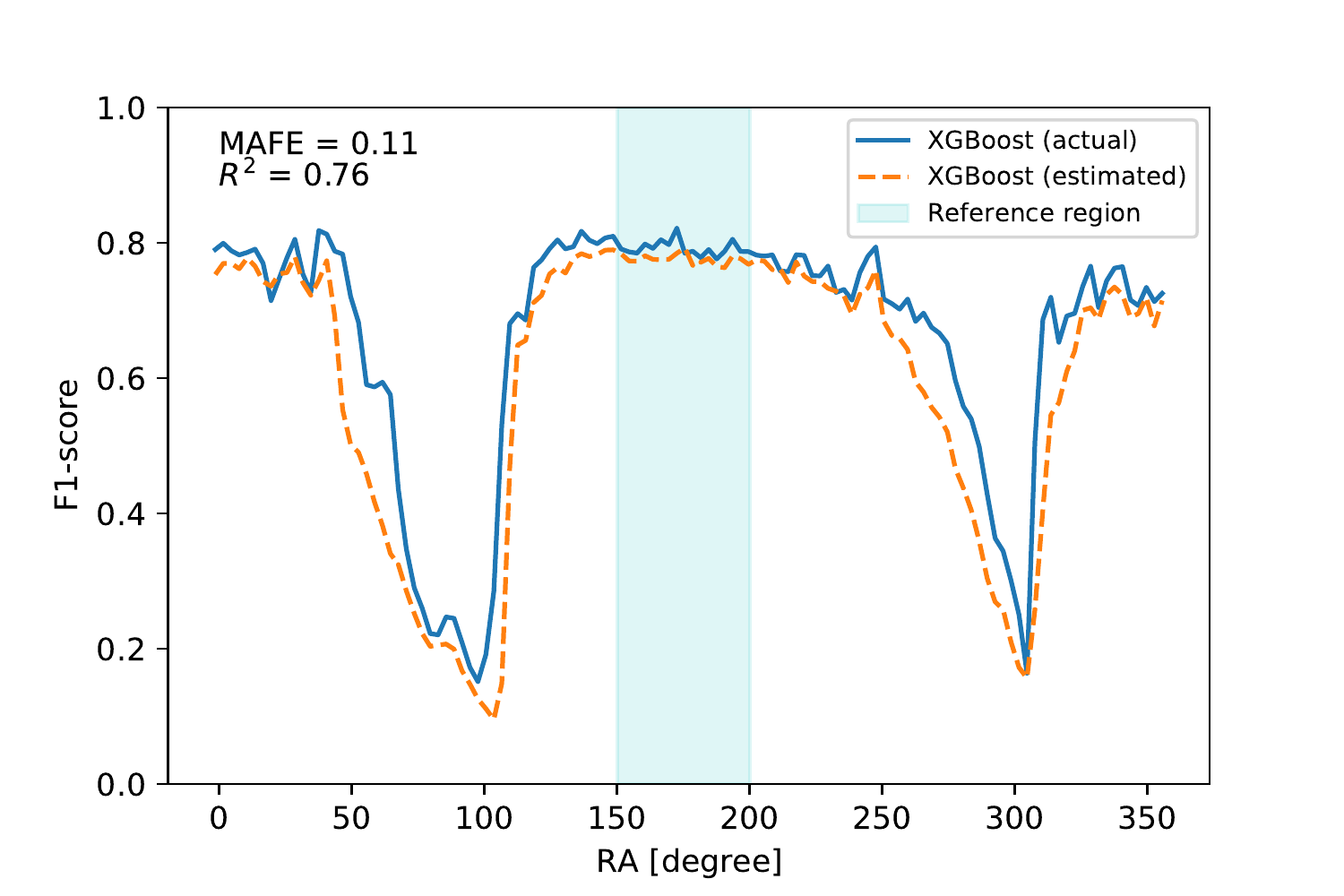}
	\includegraphics[width=\columnwidth]{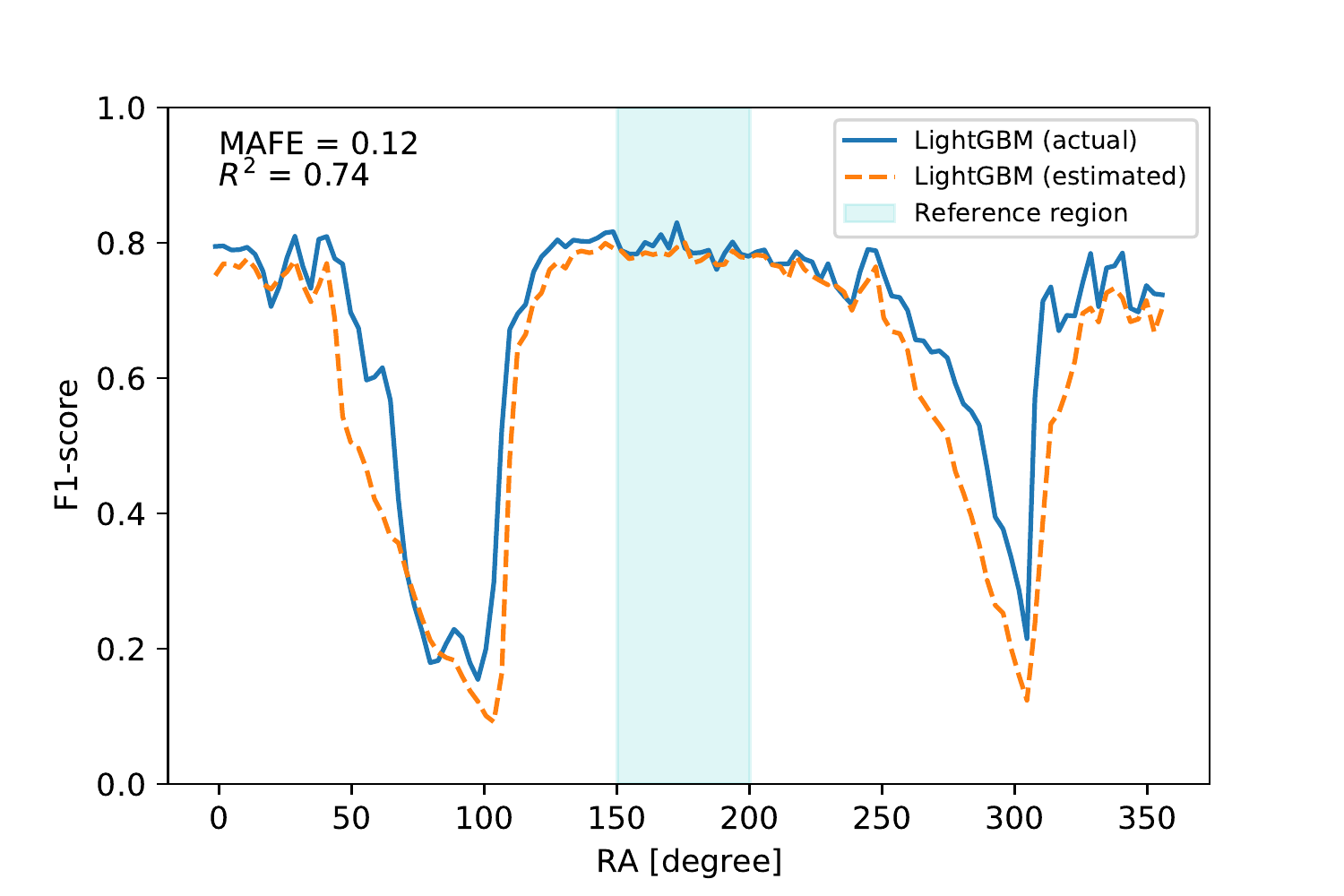}
    \caption{General characteristics of the dataset as a function of RA. 
    {\bf Top:} Estimated and actual F1-scores versus RA, for the \texttt{RandomForest} model. 
    {\bf Middle:} The same, but for the \texttt{XGBoost} model. 
    {\bf Bottom:} The same, but for the \texttt{LghtGBM} model.
    The metrics are calculated in a rolling window of length 1001 objects, with results from each window plotted at the RA value 
    that corresponds to the central element of the bin.}
   \label{fig2}
\end{figure}

For each source in the dataset that was not used for model training, the binary class and the class probability were predicted, using 
each of the three models. The F1-score versus RA is shown in Fig.~\ref{fig2} (solid blue line). The RA range from which the training sample 
was drawn is highlighted with cyan shading. 

All three models show roughtly similar behaviour of the F1-score with respect to RA: the F1-score remains near to $\sim$0.8 over most of the 
RA range, but drops dramatically to $\sim0.2$ near to ${\rm RA}\sim90$ deg and ${\rm RA}\sim310$ deg. Clearly, the population shifts seen in 
Fig.~\ref{fig1} have resulted in decreased classification performance in RA ranges where the number of stars is high relative to galaxies and
QSOs. 

Next, we apply the CBPE methodology as outlined in $\S$\ref{cbpe} to the predictions from the classifier models. The resulting F1-score estimates
are shown by the orange dashed lines in Fig.~\ref{fig2}. The estimated score follows the measured score relatively well, with $R^2$-scores of
0.92, 0.76, 0.74, and MAFE scores of 0.12, 0.11, 0.12, for the \texttt{RandomForestClassifier}, \texttt{XGBoostClassifier}, and \texttt{LightGBMClassifier} 
cases, respectively. Interestingly, the CBPE method is able to predict the dramatically low F1-scores at ${\rm RA}\sim 100$ and ${\rm RA}\sim 300$ deg, but also 
predicts the shallow gradient in the F1-score in the range $125 < {\rm RA} < 250$ deg, and many of the small fluctuations.

\section{Discussion and conclusions}
\label{conclusions}
We have introduced the NannyML CBPE method for the estimation of machine learning model performance in the absence of ground-truth labels. 
We have demonstrated that CBPE is able to predict population shift-driven changes in the performance of machine-learning models, using the 
selection of QSOs from SDSS photometry as an example use-case. 

A number of interesting implications arise from this study. Among them is the potentially game-changing possibility to 
predict the classification performance of a model in the absence of ground-truth labels, even in the presence of substantial 
population shifts between the training and test sets. Depending on the experimental set-up employed, the application of CBPE 
should allow model-selection to be performed such that models (or hyperparameters) can be chosen to maximise the expected performance 
in unlabelled (test) data, in contrast to the traditional paradigm where performance is optimized using labelled validation data. 

Another potentially fruitful use-case we envisage is the detection of population shifts in large astronomical datasets, 
and quantification of their impact on classification metrics (e.g., Fig.~\ref{fig2}). {We also argue that such analyses may be useful 
when planning wide-field surveys in which machine learning classifiers are expected to be used. For example, models trained on 
different deep, pencil-beam surveys might be evaluated on their performance over a wider survey area using CBPE, in order to select the 
model (or training sample) that is estimated to provide the highest quality result for a desired classification objective.

While the CBPE method has proven useful to predict changes in model performance under population shift, it is important to note that 
under concept drift, where changes occurs in the mapping between the features and the target, the performance estimates may be less 
accurate.

Finally, we acknowledge that the work presented in this Letter represents a restricted subset of the possible use-cases of the CBPE 
method in the field of astronomical source classification. A more detailed analysis and application to a wider range of 
classification problems is beyond the scope of this Letter, and will be the subject of a future paper.

\section*{Acknowledgements}
This work was supported by Funda\c{c}\~ao para a Ci\^encia e a Tecnologia (FCT) through grants UID/FIS/04434/2019, UIDB/04434/2020, 
UIDP/04434/2020 and PTDC/FIS-AST/29245/2017, EXPL/FIS-AST/1085/2021, and an FCT-CAPES Transnational Cooperation Project. AH also 
acknowledges support from NVIDIA in the form of a GPU under the NVIDIA Academic Hardware Grant Program.
In the development of this work, we have made use of the \texttt{Pandas} \citep{McKinney2010}, \texttt{Numpy} \citep{Harris2020}, 
\texttt{Scipy} \citep{Virtanen2020}, \texttt{Dask} \citep{Rocklin2015} and \texttt{NannyML} packages for \texttt{Python}.

\section*{Data Availability}

The Python and SQL scripts used in this study are publicly available at \href{https://github.com/humphrey-and-the-machine/f1\_prediction}
{https://github.com/humphrey-and-the-machine/f1\_prediction}.

The data were collected from the SDSS SkyServer / SciServer / CASJobs portal 
at \href{https://skyserver.sdss.org/casjobs/}{https://skyserver.sdss.org/casjobs/}.
A CSV file containing the dataset used has been made available online, hosted by MNRAS.

\bsp	
\label{lastpage}
\end{document}